\documentclass[a4paper,11pt]{article}
\usepackage{graphicx}
\usepackage{authblk}
\usepackage{amsfonts}
\usepackage{amssymb}
\usepackage{hyperref}
\begin{document}
\title{Inflation and the late time acceleration from Hossenfelder-Verlinde gravity}

\author[1,2]{Youngsub Yoon}
\author[1]{Atanu Guha}

\affil[1]{Department of Physics and Institute of Quantum Systems, \protect\\ Chungnam National University, Daejeon, 34134, Republic of Korea}
\affil[2]{Center for Theoretical Physics of the Universe, \protect\\ Institute for Basic Science (IBS), Daejeon, 34126, Republic of Korea}
\renewcommand\Affilfont{\itshape\small}

\maketitle

\begin{abstract}
We show that Hossenfelder's covariant formulation of Verlinde's emergent gravity predicts inflation and the late-time acceleration at the same time, without assuming a separate field such as inflaton, whose sole purpose is producing inflation. In particular, for the current deceleration parameter $q=-0.95$ to $-0.55$, we obtained $\lambda^2$, the mass of the imposter field, from $1.85\times 10^4$ to $2.26\times 10^4$. We also note that the value of $\lambda$ around $q=-0.93$ coincides with the inverse of fine structure constant.
\end{abstract}

\section{Introduction}
Verlinde proposed emergent gravity as an alternative to dark matter and dark energy \cite{Verlinde}. According to this theory, there is no dark matter, but our familiar Newton-Einsteinian gravity is modified. More specifically, he connected the fast galaxy rotation speed, which is currently accounted from dark matter, with our accelerating universe, currently accounted from dark energy. He showed that they do not have two separate origins. More specifically, he successfully derived the Tully-Fisher relation, an empirical relation between the asymptotic rotation speed of outermost stars in galaxies and their total baryonic mass. In particular, from Hubble's constant, he obtained the correct value of Milgrom's constant, which appears in the Tully-Fisher relation.

However, Verlinde formulated his emergent gravity only in the Newtonian limit. As a theory without a relativistic extension cannot be a complete theory, Hossenfelder proposed a covariant version of Verlinde's emergent gravity \cite{Hossenfelder}. With some elements from Verlinde's emergent gravity, she proposed a new Lagrangian that mimics the behavior of Verlinde's emergent gravity. Besides the advantage that the relativistic extension is possible, Hossenfelder noted that a Lagrangian formulation will be free of theoretical inconsistencies that plague Verlinde's original formulation of emergent gravity. Perhaps due to these theoretical weaknesses of Verlinde's formulation of emergent gravity, one of us \cite{YoonHossenfelder} surprisingly found that Hossenfelder's version of Verlinde's emergent gravity does not reduce to Verlinde's original formulation of emergent gravity but to MOND (Modified Newtonian Dynamics) proposed by Milgrom \cite{MOND1, MOND2} in 1983 to explain the Tully-Fisher relation.

As Hossenfelder's formulation of Verlinde's emergent gravity passed the test in the galactic scale \cite{Hossenfelder:2018vfs}, we need to test it further in the cosmic scale. In this paper, we perform the first such test by calculating the evolution of the scale factor, i.e., the expansion of our universe in the history of universe.

In particular, we show that both inflation \cite{Guth:1980zm} and the current acceleration can be explained by Hossenfelder-Verlinde theory of gravity, without assuming a new field such as inflaton, whose only purpose is producing inflation; in Hossenfelder's covariant formulation of Verlinde's emergent gravity, there is what she called the ``imposter field,'' which, as we will show, can produce both inflation and the late-time acceleration. The cosmological constant is also not necessary.

The organization of this paper is as follows. In Section \ref{covarianthossenfelder}, we will present the covariant formulation of Verlinde gravity due to Hossenfelder. In Section \ref{Einstein}, we will obtain Einstein's equation for Hossenfelder-Verlinde gravity. In Section \ref{results}, we present our simulation results. In Section \ref{discussion}, we conclude our paper.

\section{Covariant formulation of Verlinde gravity}\label{covarianthossenfelder}
In this section, we review Hossenfelder's work \cite{Hossenfelder}. To introduce the covariant Lagrangian for Verlinde's emergent gravity, it is necessary to define the elastic strain tensor. It is given in terms of $u_\mu$, called the displacement field or the imposter field, as follows. The signature of the metric is $(-,+,+,+)$.

\begin{equation}
\epsilon_{\mu\nu}=\nabla_\mu u_\nu+\nabla_\nu u_\mu
\end{equation}

We also have
\begin{equation}
u=\sqrt{-u^\mu u_\mu},\qquad \phi\equiv \frac uL \qquad \epsilon\equiv\epsilon^\mu_\mu,\qquad n^\mu\equiv\frac{u^\mu}{u}
\end{equation}
and
\begin{equation}
\chi=-\frac 14 \epsilon_{\mu\nu}\epsilon^{\mu\nu}+\frac 13 \epsilon^2
\end{equation}

The total Lagrangian is given by$\footnote{As noted in \cite{YoonHossenfelder}, we fix the sign for $\mathcal L_{int}$ in (\ref{Lint}). Otherwise, $\rho_{int}$ in Friedmann's equation $H^2=(8 \pi G/3)(\rho_s+\rho_M+\rho_{int})$ becomes negative, and we don't obtain the correct MOND equation.}$
\begin{equation}
\mathcal L_{tot}=\frac  12 m_p^2\mathcal R +\mathcal L_M+\mathcal L_{int}+\mathcal L_{s}
\end{equation}
where $m_p^2=1/(8\pi G)$
\begin{equation}
\mathcal L_{int}=\frac 1L u^\mu n^\nu T_{\mu\nu}=\frac{u^\mu u^\nu}{Lu}T_{\mu\nu}\label{Lint}
\end{equation}
\begin{equation}
\mathcal L_{s}=\frac{m_p^2}{L^2}\chi^{3/2}-\frac{\lambda^2 m_p^2}{L^4} u_{\kappa}u^{\kappa}\label{Ls}
\end{equation}
where $\mathcal L_M$ is the visible matter Lagrangian and $L$ gives the length scale of our Verlinde's de Sitter universe. $\lambda$ is not a priori known.  

\section{Einstein's equation}\label{Einstein}
In this section, by closely following \cite{Comment}, we review Einstein's equation for the FLRW universe in case of the covariant formulation of Verlinde's emergent gravity.$\footnote{Unlike \cite{Comment}, we didn't impose $\xi=0$ in the limit when $t\rightarrow \infty$. Instead, we fixed $N=1$ for (15) in \cite{Comment}.}$ As in \cite{Hossenfelder} and \cite{Comment}, we assume that the spatial part of the imposter field $u_\mu$ is zero.

\begin{equation}
T_{\mu\nu}=(T_s)_{\mu\nu}+(T_M)_{\mu\nu}+(T_{int})_{\mu\nu}
\end{equation}
\begin{equation}
(T_s)_{\mu\nu}=\frac{m_p^2}{L^2}\sqrt{\chi}\left(\frac 32 \epsilon_{\mu\alpha}\epsilon^{\alpha}_{\nu}-2\epsilon_{\mu\nu}\epsilon+\chi g_{\mu\nu}\right)+\frac{\lambda^2 m_p^2}{L^4}(2u_\mu u_\nu +g_{\mu\nu}u^2)
\end{equation}
\begin{equation}
(T_{int})_{\mu\nu}=-\frac{4u_\mu u^\gamma (T_M)_{\nu\gamma}}{Lu}-\frac{u_\nu u_\mu u^\kappa u^\gamma (T_M)_{\kappa\gamma}}{Lu^3}+g_{\mu\nu}\frac{u^\kappa u^\gamma (T_M)_{\kappa\gamma}}{Lu}
\end{equation}
\begin{equation}
	ds^2=-dt^2+e^{2v(t)} (dr^2+r^2(d\theta^2+\sin^2\theta d\phi^2))
\end{equation}
\begin{equation}
	u_t=L e^{2\xi(t)},\quad u_r=u_\theta=u_\phi=0
\end{equation}
\begin{equation}
	\epsilon_t^t=-4L e^{2\xi}\dot\xi,\quad \epsilon_r^r=\epsilon_\theta^\theta=\epsilon^\phi_\phi=-2L e^{2\xi}\dot{v}
\end{equation}
\begin{equation}
	\chi=L^2 e^{4\xi}\left(\frac 43 \dot\xi^2+16\dot\xi\dot v+9\dot v^2\right)
\end{equation}
\begin{eqnarray}
	&&(T_s)^t_t=\frac{m_p^2}{3}  e^{4\xi}\sqrt{\chi}(-20\dot \xi^2-96\dot\xi\dot v+27\dot v^2)- \frac{\lambda^2}{L^2} m_p^2 e^{4\xi}\\
	&&(T_s)^r_r=(T_s)^\theta_\theta=(T_s)^\phi_\phi=\frac{m_p^2}{3} e^{4\xi}\sqrt{\chi}(4\dot\xi^2-27\dot v^2)+ \frac{\lambda^2}{L^2} m_p^2 e^{4\xi}\\
	&&(T_M)_{tt}=\rho_{b}+\rho_{r},\qquad (T_M)^t_t=-(T_M)_{tt},\qquad (T_M)^r_r=\frac{\rho_{r}}{3}\\
	&&(T_{int})^t_t=-2 e^{2\xi}(T_M)_{tt},\qquad (T_{int})^r_r=e^{2\xi}(T_M)_{tt}\\
	&&\frac{\partial}{\partial t}\left(\rho_b(1+2 e^{2\xi})\right)+3\dot v \rho_b(1+3e^{2\xi})=0 \label{rhob}\\
	&&\frac{\partial}{\partial t}\left(\rho_r(1+2 e^{2\xi})\right)+\dot v \rho_r(4+9e^{2\xi})=0   \label{rhor}
\end{eqnarray}
where the last two equations come from the conservation of energy-momentum tensor.

The Einstein tensor is given by
\begin{eqnarray}
	&&- 3m_p^2 \dot v^2=G_0^0\\
	&&-m_p^2 (2\ddot v+3 \dot v^2)=G_r^r=G_\theta^\theta=G_\phi^\phi
\end{eqnarray}

We solved the two independent Einstein equations numerically (i.e., $G^0_0=T^0_0$ and $G^r_r=T^r_r$) and two energy-momentum tensor conservation equations (i.e., (\ref{rhob}) and (\ref{rhor})) using Mathematica. They are four coupled differential equations. We will discuss our results in the next section.

\section{Our results}\label{results}
For the Hubble constant, we used $H_0=73 \mathrm{km/s/Mpc}$. Also, $L$ can be expressed by Milgrom's constant $a_M$ by the following equation \cite{YoonHossenfelder}
\begin{equation}
	L=\frac{4\sqrt 2}{3a_M} c^2=(2.33\pm 0.03)\times 10^5~\mathrm{Myr}\label{aM}
\end{equation}
where we used $a_M=(0.77\pm 0.01)\times 10^{-10}\mathrm{m/s^2}$ \cite{Hossenfelder:2018vfs}. For the baryonic density and the radiation density, we used $\Omega_b h^2=0.02237\pm 0.00015$ and $\Omega_r h^2=2.47\times 10^{-5}$ \cite{Planck:2018vyg}.

We gave the current scale factor, i.e., $a(0)=1$ ($v(0)=0$), the current Hubble constant, the current acceleration, the current baryonic density and the current radiation density as the boundary conditions for the differential equations. While the Hubble constant at the present universe can be quite precisely measured there is a large observational uncertainty for the current acceleration of universe, which is often parameterized by the deceleration parameter $q$. Moreover, depending on which model of universe one uses to fit the observed expansion of our universe, $q$ varies greatly. Therefore, we tried various values of $q$: $-0.55, -0.70, -0.85, -0.95$. For each $q$, we ran simulations for various $\lambda^2$.

\begin{figure}
	\centering
	\includegraphics[height=60mm]{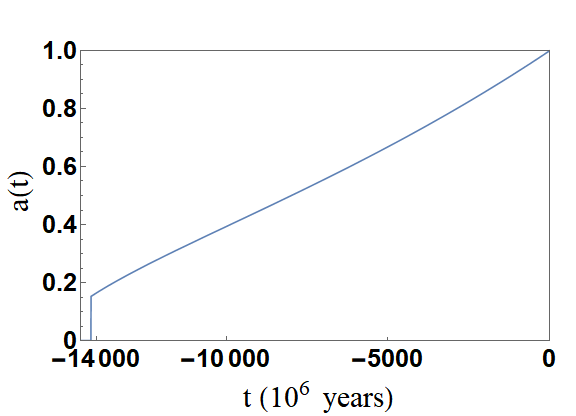}
	
	\caption{The evolution of scale factor for $q=-0.70$ and $\lambda^2=20000$. We see that the scale factor suddenly drops to zero at around $a=\mathcal O(0.1)$ as we go back in time.}
	\label{070himanshu20000}
\end{figure}

Interestingly, we found out that for most values of $\lambda^2$, the numerical solutions to differential equations stop around $a=\mathcal O(0.1)$ and drop to $a=0$ vertically. See Fig. \ref{070himanshu20000} for an example. In other words, for most values of $\lambda^2$, universe expanded from $a=0$ to $a=\mathcal O(0.1)$ very rapidly, then the expansion rate became suddenly moderate. This behavior is expected from inflation except for the fact that the scale factor at the exit is too large.

The scale factor at the inflation exit can never be as big as $\mathcal O(0.1)$. Big bang nucleosynthesis \cite{Cyburt:2015mya} is a very successful theory that correctly predicts the abundance of light elements in our current universe. If the inflation exited too late, all the delicate predictions of big bang nucleosynthesis would be rendered invalid. It also goes without saying that the recombination, which is responsible for the accurately measured CMB anisotropy happened at a scale factor $\mathcal O(0.001)$, which is far smaller than $\mathcal O(0.1)$. Therefore, we fine-tuned $\lambda^2$ that yields a small scale factor when it vertically drops in the graph. 

\begin{figure}[!ht]
	\centering
	\begin{minipage}[t]{.48\textwidth}
		\centering
		\includegraphics[width=\textwidth]{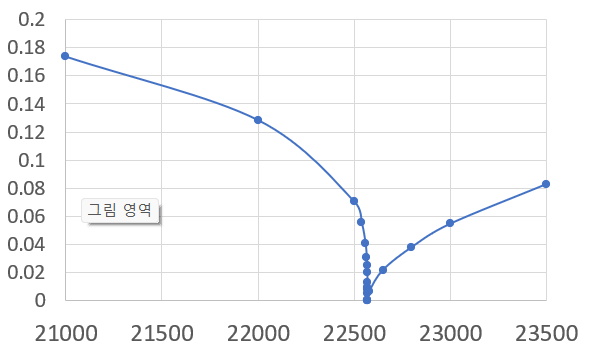}
		\caption{the initial scale factor as a function of $\lambda^2$. Here, we have $q=-0.55$.}
		\label{055}
	\end{minipage}%
	\hfill
	\begin{minipage}[t]{.48\textwidth}
		\centering
		\includegraphics[width=\textwidth]{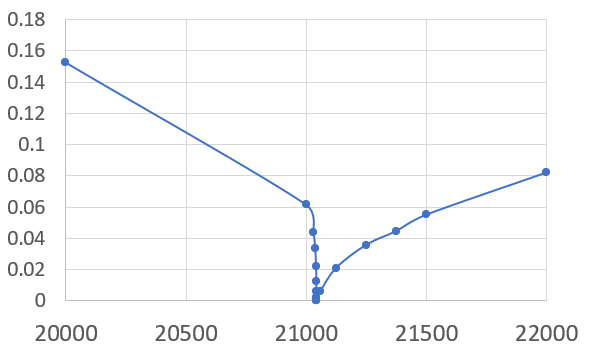}
		\caption{the initial scale factor as a function of $\lambda^2$. Here, we have $q=-0.70$.}
		\label{070}
	\end{minipage}
\end{figure}

\begin{figure}[!ht]
	\centering
	\begin{minipage}[t]{.48\textwidth}
		\centering
		\includegraphics[width=\textwidth]{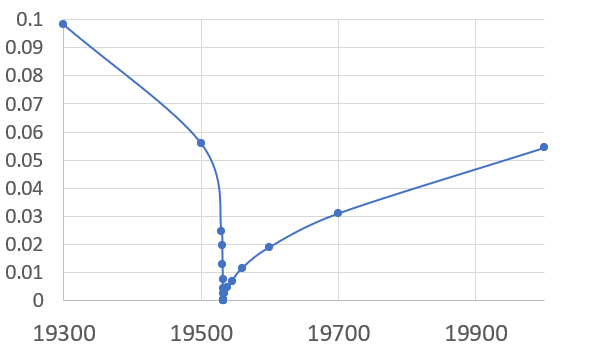}
		\caption{the initial scale factor as a function of $\lambda^2$. Here, we have $q=-0.85$.}
		\label{085}
	\end{minipage}%
	\hfill
	\begin{minipage}[t]{.48\textwidth}
		\centering
		\includegraphics[width=\textwidth]{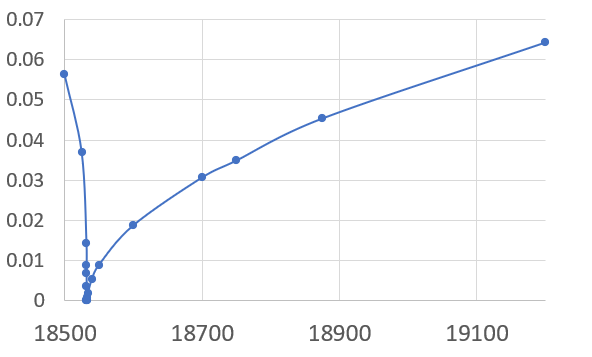}
		\caption{the initial scale factor as a function of $\lambda^2$. Here, we have $q=-0.95$.}
		\label{095}
	\end{minipage}
\end{figure}

We listed the initial scale factor as a function of $\lambda^2$ in Tables \ref{-0.55}, \ref{-0.70}, \ref{-0.85} and \ref{-0.95}. In case where the initial scale factor was too small to be noticed by Mathematica computation, we wrote ``0'' for the initial scale factor. Such are the most realistic cases for our Universe. For the graphic representations of these tables, see Fig. \ref{055}, \ref{070}, \ref{085} and \ref{095}. Based on these tables, in Table \ref{finalresults}, we listed the values of $\lambda^2$ a sensible universe (i.e., inflation exit at a small scale factor) needs to have, given its current deceleration parameter (i.e., acceleration).

\begin{figure}[!ht]
	\centering
	\includegraphics[height=60mm]{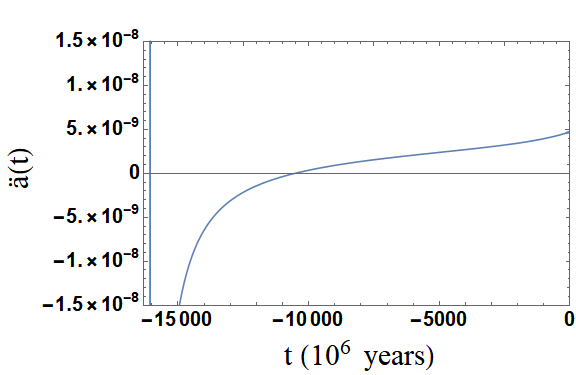}
	\caption{The acceleration has a high peak in the beginning of universe. $q=-0.85$, $\lambda^2=19532$}
	\label{085himanshu19532}
\end{figure}

The inflation-like behavior is also apparent when we plot the acceleration as a function of time. See Fig. \ref{085himanshu19532} for an example. There is a peak at the beginning of the universe. However, when we tried to capture the exact moment of inflation, we faced obstacles due to the computing inaccuracy of Mathematica. When we tried to zoom into the beginning of universe, the numerical solution to the differential equations was very unstable. We tried to increase the working precision and decrease the step size, but the problem persisted.


\section{Discussions and Conclusions}\label{discussion}
In this paper, we successfully showed that Hossenfelder-Verlinde gravity can explain both inflation and the late-time acceleration. However, as we mentioned earlier, there was a computing difficulty for Mathematica to capture the exact moment of inflation, which must be overcome by subsequent research. 

We also saw that the condition that the exit of inflation happens at sufficiently low scale factor determines $\lambda^2$ if the current deceleration parameter of our universe is known. An independent method to check it would be obtaining the $\lambda^2$ and the current deceleration parameter by fitting supernovae data with Hossenfelder-Verlinde gravity in the manner of \cite{Bouali:2023rdx}. We determined $\lambda^2$ from the early time information about our Universe. The authors of \cite{Bouali:2023rdx} determined various parameters from the late time information about our Universe. Therefore, if a test is performed in a similar manner, it will a good consistency check. Future research also needs to address whether Hossenfelder-Verlinde gravity can explain the Hubble tension and the CMB anisotropy spectrum.

Finally, we would like to mention an interesting coincidence we found. Our results show that the value of $\lambda$ for $q=-0.93$ is close to the value of the inverse of the fine structure constant. If the deceleration parameter turns out to be about $q=-0.93$, it would need a further investigation why such a coincidence happens. 

However, it may be possible that the deceleration parameter is somewhat different from this value, even though $\lambda$ is exactly the inverse of the fine structure constant, considering that the value of $L$ we used has an error. $L$ is directly related to $a_M$, which is obtained in \cite{Hossenfelder:2018vfs}. The error of $a_M$ is about 1\%, but as mentioned in \cite{Hossenfelder:2018vfs}, the actual error can be bigger, as the 1\% error is only the one of the statistical fit, which is quite small due to the large number of data; in reality, the normalization of the stellar mass-to-light ratio, which is crucial in determining $a_M$ has about 30 \% error.

\section*{Acknowledgement}
We thank Pablo Soler Gomis, Boris Latosh, Takahiro Terada and Tae Hyun Jung for helpful discussions. This work is supported by the National Research Foundation of Korea [NRF-2019R1C1C1005073(YY, AG) and NRF-2021R1A4A2001897(YY)] and by IBS under the project code, IBS-R018-D1 (YY).

\begin{table}
	\centering
	\begin{tabular}{|l|l|}
		\hline
		$q$ & $\lambda^2$  \\\hline
		-0.55 & 22569.76  \\
		-0.70 & 21043.14  \\
		-0.85 & 19531.824 \\
		-0.95 & 18532.253  \\
		\hline
	\end{tabular}
	\caption{$\lambda^2$ as a function of deceleration parameter $q$ for sensible universe}
	\label{finalresults}
\end{table}

\begin{table}
	\centering
	\begin{tabular}{|l|l|l|l|l|l|}
		\hline
		$\lambda^2$ & initial $a$ & $\lambda^2$ & initial $a$ & $\lambda^2$ & initial $a$ \\\hline
		21000 & 0.1735& 22569  & 0.0198 & 22569.8 & 0.00012\\
		22000 & 0.1284& 22569.6& 0.0127 & 22570 & 0.00048\\
		22500 & 0.0706& 22569.7& 0.0095 & 22580&0.0068\\
		22540 & 0.0554& 22569.73& 0.0077&22650&0.0218\\
		22560 & 0.0405& 22569.75& 0.0052&22800&0.038 \\
		22566 & 0.031 & 22569.76& 2.7$\times 10^{-5}$& 23000&0.0546\\
		22568 & 0.025 & 22569.77& 7.9$\times 10^{-5}$ & 23500 &0.0828\\
		\hline
	\end{tabular}
	\caption{the initial scale factor as a function of $\lambda^2$. The deceleration parameter $q$ is $-0.55$.}
	\label{-0.55}
\end{table}

\begin{table}
	\centering
	\begin{tabular}{|l|l|l|l|l|l|}
		\hline
		$\lambda^2$ & initial $a$ & $\lambda^2$ & initial $a$ & $\lambda^2$ & initial $a$ \\\hline
		20000 & 0.1525& 21043.14  & 0 & 21060 & 0.006\\
		21000 & 0.0614& 21043.15& 2.3$\times 10^{-5}$ & 21125 & 0.021\\
		21030 & 0.0439& 21043.19& 6.9$\times 10^{-5}$ & 21250&0.036\\
		21038 & 0.0337& 21043.25& 0.0003 &21375&0.044\\
		21042 & 0.0221& 21043.5& 0.0008&21500&0.055 \\
		21043 & 0.0123 & 21044& 0.0010& 22000&0.082\\
		21043.13 & 0.0058 & 21045& 0.0022 &  &\\
		\hline
	\end{tabular}
	\caption{the initial scale factor as a function of $\lambda^2$. The deceleration parameter $q$ is $-0.70$.}
	\label{-0.70}
\end{table}

\begin{table}
	\centering
	\begin{tabular}{|l|l|l|l|l|l|}
		\hline
		$\lambda^2$ & initial $a$ & $\lambda^2$ & initial $a$ & $\lambda^2$ & initial $a$ \\\hline
		19300 & 0.0984 & 19531.823  & 0.0027 & 19534 & 0.002\\
		19500 & 0.0561 & 19531.824 & 9.1$\times 10^{-6}$ & 19538 & 0.005\\
		19530 & 0.0248 & 19531.825& 1.9$\times 10^{-5}$ & 19545&0.007\\
		19531 & 0.0195 & 19531.83& 4.6$\times 10^{-5}$ &19560&0.012\\
		19531.6 & 0.0128 & 19531.85 & 0.00012 &19600&0.019 \\
		19531.8 & 0.0076 & 19531.9 & 0.00026 & 19700&0.031\\
		19531.82 & 0.0045 & 19532 & 0.00052 & 20000 & 0.054 \\
		\hline
	\end{tabular}
	\caption{the initial scale factor as a function of $\lambda^2$. The deceleration parameter $q$ is $-0.85$.}
	\label{-0.85}
\end{table}

\begin{table}
	\centering
	\begin{tabular}{|l|l|l|l|l|l|}
		\hline
		$\lambda^2$ & initial $a$ & $\lambda^2$ & initial $a$ & $\lambda^2$ & initial $a$ \\\hline
		18500 & 0.0563 & 18532.255  & 5.1$\times 10^{-5}$ & 18550 & 0.009\\
		18525 & 0.0369 & 18532.26 & 7.9$\times 10^{-5}$ & 18600 & 0.019\\
		18532 & 0.0143 & 18532.3& 9.0$\times 10^{-5}$ & 18700 &0.031\\
		18532.2 & 0.0089 & 18532.5& 0.00034 &18750&0.035\\
		18532.23 & 0.0067 & 18533 & 0.00092 &18875&0.045 \\
		18532.245 & 0.0036 & 18534 & 0.00178 & 19200&0.064\\
		18532.253 & 0 & 18540 & 0.0054 &  &  \\
		\hline
	\end{tabular}
	\caption{the initial scale factor as a function of $\lambda^2$. The deceleration parameter $q$ is $-0.95$.}
	\label{-0.95}
\end{table}

\end{document}